\documentclass[a4paper,10pt]{article}

\usepackage[super,compress]{cite}
\usepackage{graphicx}
\usepackage{amsmath,amssymb}

\def\units#1{\hbox{$\,{\rm #1}$}}
\def\degrees{\hbox{$^\circ$}}

\begin{document}
\markboth{M.~N.~MAZZIOTTA}
{Indirect searches for dark matter with the Fermi LAT instrument}

\title{Indirect searches for dark matter with the Fermi LAT instrument}

\author{M.~N.~MAZZIOTTA \\ on behalf of the Fermi-LAT Collaboration \\ 
\\
Istituto Nazionale di Fisica Nucleare, Sezione di Bari, 70125 Bari, Italy \\
\texttt{mazziotta@ba.infn.it} \\
\\
Proceeding of the 2013 Lepton Photon Conference}

\maketitle


\begin{abstract}
In this review the current status of several searches for particle dark matter with 
the Fermi Large Area Telescope instrument is presented. In particular, the current limits 
on the weakly interacting massive particles, obtained from the analyses of 
gamma-rays and cosmic-ray electron/positron data, will be illustrated.

\end{abstract}



\section{Introduction}
\label{sec:intro}
	
A wide range of cosmological observations in the Universe, including large scale structures, 
the cosmic microwave background and the isotopic abundances resulting from the primordial 
nucleosynthesis, provide an evidence for a sizable non-baryonic and extremely-weakly 
interacting cold dark matter (DM)\cite{Roos:2012cc}.

Weakly interacting massive particles (WIMPs) provide a theoretically appealing class 
of candidates for the obscure nature of DM\cite{Jungman:1995df,Bergstrom:2000pn,Bertone:2004pz}, 
with the lightest supersymmetric neutralino ($\chi$) often taken as a useful template for such 
a WIMP. It is often argued that the thermal production of WIMPs in the early universe 
generically leads to a relic density that coincides with the observed order of magnitude 
of DM fraction on cosmological scales, $\Omega_\chi= 0.229 \pm 0.015$, to the total 
energy density of the universe\cite{wmap}.

The indirect search for DM is one of the main items in the broad Fermi Large Area 
Telescope (LAT)\cite{Atwood2009,Abdo:2009gy,Ackermann:2012kna} Science menu~\cite{balt}.
It is complementary to direct searches for the recoil of WIMPs off the nuclei being 
carried out in underground facilities~\cite{DRUKIER:2013lva}
and at collider experiments searches for missing 
transverse energy. In this review the word ``indirect'' denotes search for a signature 
of WIMP annihilation or decay processes through the final products (gamma rays, 
electrons and positrons) of such processes. 

Among possible messengers for such indirect searches, gamma rays play a pronounced 
role as they propagate essentially unperturbed through the Galaxy and therefore 
directly point to their sources, leading to distinctive spatial signatures; 
an even more important aspect, as we will see, is the appearance 
of pronounced spectral signatures.  

Among many other ground-based and space-borne instruments, the LAT plays a prominent role 
in this search through a variety of distinct search targets: gamma-ray lines, Galactic 
and isotropic diffuse gamma-ray emission, dwarf satellites, cosmic ray (CR) 
electrons and positrons.

\section{The LAT instrument}
\label{sec:lat}

The LAT is a pair-conversion gamma-ray telescope designed to measure gamma rays in the 
energy range from $20\units{MeV}$ to more than $300\units{GeV}$. In this paper a brief 
description of the LAT is given, while full details can be found 
in\cite{Atwood2009,Abdo:2009gy,Ackermann:2012kna}.

The LAT is composed of a $4 \times 4$ array of $16$ identical towers designed to 
convert incident gamma rays into $e^{+} e^{-}$ pairs, and to determine their arrival 
directions and energies. Each tower hosts a tracker module and a calorimeter module. 
Each tracker module consists of $18$ x-y planes of silicon-strip detectors, interleaved 
with tungsten converter foils, for a total on-axis thickness equivalent to $1.5$ radiation 
lengths (r.l.). Each calorimeter module, $8.6$ r.l. on-axis thick, hosts $96$ CsI(Tl) 
crystals, hodoscopically arranged in $8$ perpendicular layers. The instrument is 
surrounded by a segmented anti-coincidence detector that tags the
majority of the charged-particle background. 

The field of view is $\sim 2.4~sr$ and LAT observes the entire sky about 
every 3 hours (2 orbits). These features make the LAT a highly-sensitive instrument for dark matter 
searches. The operation of the instrument during the first five years of the mission was smooth. 
The LAT has been collecting science data for more than 99\% of the time spent 
outside the South Atlantic Anomaly (SAA). The remaining tiny fractional down-time 
accounts for both hardware issues and detector calibrations\cite{Abdo:2009gy,Ackermann:2012kna}.

Over the first five years of mission the LAT collaboration has put a considerable
effort toward a better understanding of the instrument and of the environment in which
it operates. In addition to that a continuous effort was made in order to make the
advances public as soon as possible. In August 2011 the first new event classification
(Pass-7) since launch was released, along with the corresponding Instrument Response
Functions. Later, in fall 2013, a new version of Pass-7 event classification was 
released (Pass-7 reprocessed) with improved calibrations for the light yield and 
asymmetry in the calorimeter crystals\cite{pass7rep}. The new calorimeter calibrations 
improve the in-flight point-spread function (PSF) above $\sim 3$ GeV and correct for 
the small ($\sim$ 1\% per year), expected degradation in the light yield of the 
calorimeter crystals measured in flight data. Consequently, the absolute energy 
scale has shifted upwards by a few percent in an energy and in a time-dependent 
manner. In addition, the re-calibration of the calorimeter light asymmetry leads 
to a statistical re-shuffling of the events classiffied as photons.

\section{Indirect Dark Matter searches with gamma-rays}
\label{sec:dmgray}

The high-energy gamma-ray sky is dominated by diffuse emission: more than 70\% of 
the photons detected by the LAT are produced in the interstellar space of our Galaxy 
by interactions of high-energy cosmic rays with matter and low-energy radiation fields. 
An additional diffuse component with an almost-isotropic distribution (and therefore 
thought to be extragalactic in origin) accounts for another significant fraction of 
the LAT photon sample. The rest consists of various different types of point-like 
or extended sources: Active Galactic Nuclei (AGN) and normal galaxies, pulsars 
and their relativistic wind nebulae, globular clusters, binary systems, shock-waves 
remaining from supernova explosions and nearby solar-system bodies like the Sun and the Moon.

\begin{figure}[!t]
\centerline{\includegraphics[width=0.8\textwidth]{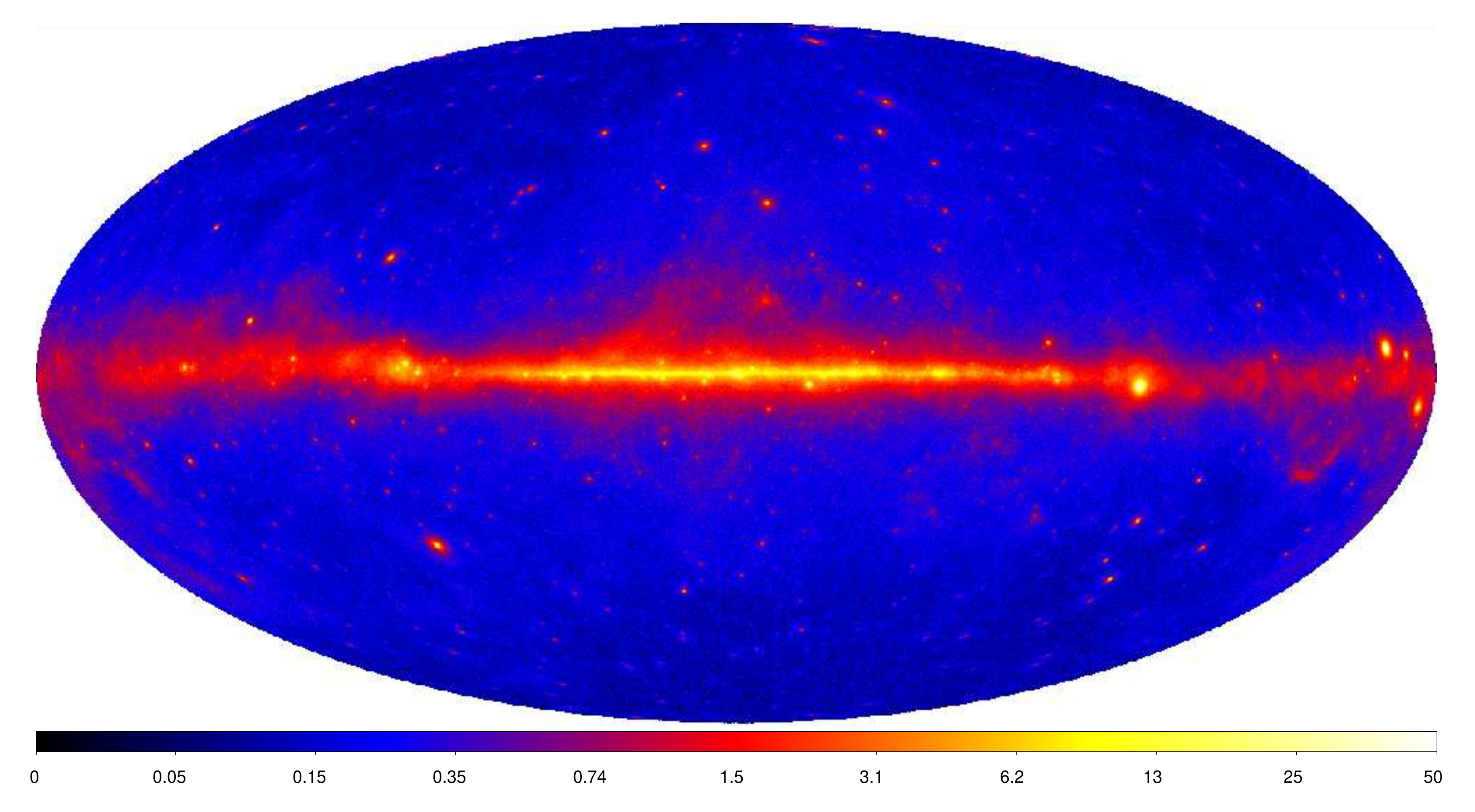}}
\caption{Sky map of the energy flux derived from the LAT data taken in the first 2 years of observation, Aitoff projection in Galactic coordinates. The image shows the gamma-ray energy flux for energies between 100~MeV and 10~GeV, in units of 10$^{-7}$ erg cm$^{-2}$ s$^{-1}$ sr$^{-1}$.  Reprinted from Ref.~\citen{2fgl}.}
\label{fig1}
\end{figure}

The Second Fermi-LAT catalog (2FGL)\cite{2fgl} contains 1873 sources detected 
and characterized in the energy range from 100 MeV to 100 GeV. Amonge these sources,
127 are considered as being firmly identified and 1171 as being reliably associated 
with counterparts of known or likely gamma-ray-producing source classes. 
A sky map of the energy flux derived from the first two years of LAT data is shown in Fig.~\ref{fig1}.

The peculiar morphology of annihilation signals, tracing directly the DM density, offers 
a convenient handle for discriminating signals from backgrounds (foregrounds). The DM detectability 
of any particular region in the Universe strongly depends on the density distribution of DM particles
along the line of sight. The main foreground for the DM search is due to the 
standard astrophysical origin of $\gamma$ emission. The most relevant targets are the 
Galactic center (GC), the Galactic halo, the dwarf spheroidal galaxies and the galaxy clusters. 
Other important targets are DM clumps and the Sun. Further information can be obtained by
studying the angular power spectrum of the isotropic gamma-ray background (IGRB).

Gamma rays can be produced by dark matter annihilations in two major ways: 
\begin{itemize}
 \item annihilation into other particles, which eventually produces gamma rays either through 
pion production, or final state bremsstrahlung and inverse Compton radiation from leptonic channels;
 \item direct annihilation to $\gamma$X, where X usually is another neutral state, 
typically a $\gamma$-ray, a Z or a Higgs boson.
\end{itemize}
The first process will result into a continuous gamma-ray energy spectrum, 
while the second one will result into a line in the gamma-ray spectrum

The expected DM-induced differential gamma-ray flux from annihilation $\Phi_{\gamma}(E, \Delta \Omega)$ 
at a given energy $E$, (in units of $\units{photons~cm^{-2}~s^{-1}~GeV^{-1}}$) from WIMP annihilation 
in a region covering a solid angle $\Delta \Omega$ and centered on a DM source, 
can be factorized as~\cite{Mazziotta:2012ux}:

\begin{equation}
\Phi_{\gamma}(E, \Delta \Omega) = J(\Delta \Omega) \times \Phi^{PP}(E)
\label{eq:DMflux}
\end{equation}
where $J(\Delta \Omega)$ (in units of $\units{GeV^{2}~cm^{-5}~sr}$) is the 
{\em ``astrophysical factor''} or {\em ``J-factor''}, i.e. the line of sight (l.o.s.) integral of the 
DM density $\rho$ (in units of $\units{GeV~cm^{-3}}$) squared in the direction of observation $\psi$ over the 
solid angle $\Delta \Omega$:

\begin{equation}
J(\Delta \Omega) = \int_{\Delta \Omega} d\psi \int_{l.o.s.} dl \rho^{2}(l, \psi)
\label{eq:jfactor}
\end{equation}

The term $\Phi^{PP}(E)$ (in units of $\units{GeV^{-3}~cm^{3}~s^{-1}~sr^{-1}}$) is the {\em ``particle physics factor''},
that encodes the particle physics properties of the DM, and for a given WIMP mass $m_{\chi}$ is given by: 

\begin{equation}
\Phi^{PP}(E) =  
\frac{1}{4 \pi} \frac{\langle \sigma v \rangle}{2 m^{2}_{\chi}} \sum_{f} N_{f}(E, m_{\chi}) B_{f}
\label{eq:PPfactor}
\end{equation}
where $\langle \sigma v \rangle$ is the average velocity weighted annihilation cross section of 
the two annihilating particles (for which we assume $\bar{\chi}=\chi$), while $B_f$ 
and $N_{f}(E, m_{\chi})$ are respectively the branching ratio and the differential photon 
spectrum of each pair annihilation final state $f$ (i.e. the number of photons per annihilation). 
An often quoted reference value for $\langle \sigma v \rangle$ is the so-called 
{\em ``thermal cross section''} $\langle \sigma v \rangle = 3 \times 10^{-26} cm^3 s^{-1}$, 
which is the annihilation rate expected for thermally produced WIMPs in simplest case 
(i.e. s-wave annihilation without resonances or co-annihilations\cite{griest}).

In case of DM decay the corresponding production spectrum is obtained by replacing 
$\langle \sigma v \rangle /2 m^{2}_{\chi}$ with $\Gamma/m_\chi$, where $\Gamma$ is the 
decay rate, in Eq.~\ref{eq:PPfactor}, and $\rho^{2} \rightarrow \rho$ in Eq.~\ref{eq:jfactor}.

\begin{figure*}[!t]
\begin{center}
\includegraphics[width=0.45\textwidth]{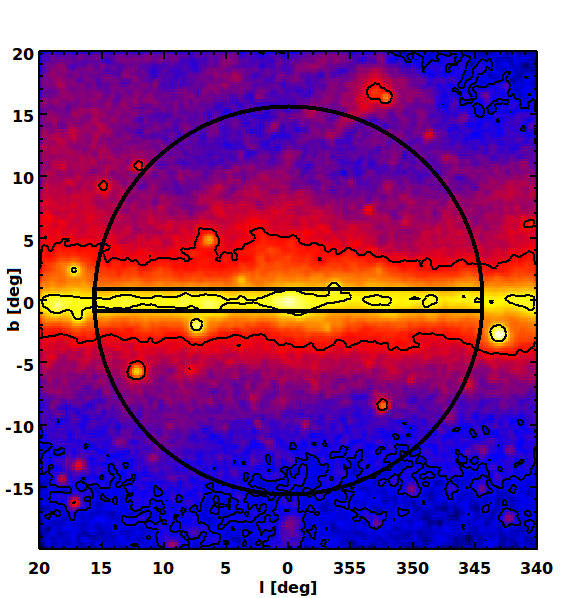}
\includegraphics[width=0.45\textwidth]{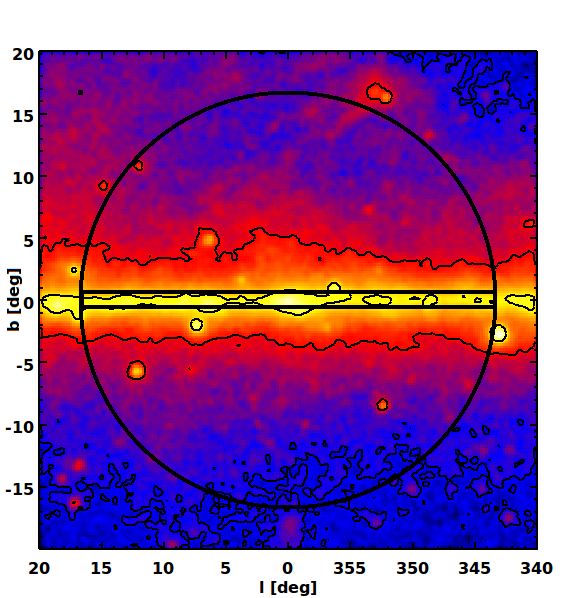}
\includegraphics[width=0.45\textwidth]{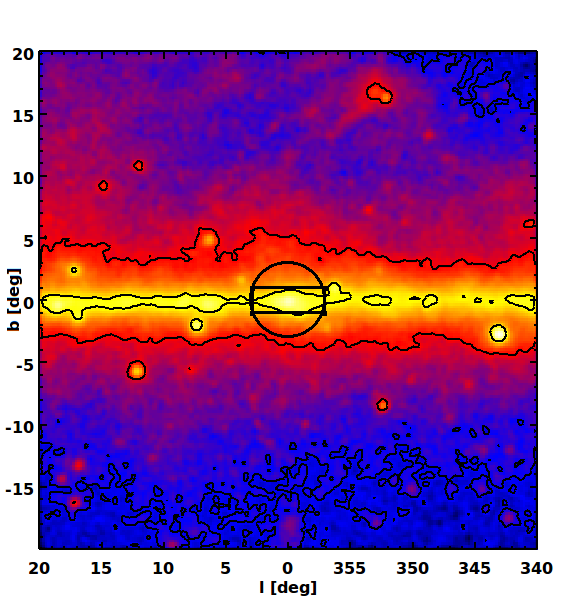}
\includegraphics[width=0.45\textwidth]{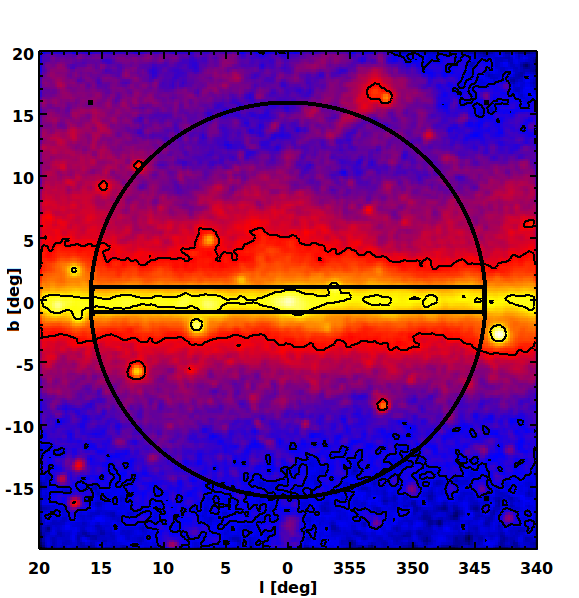}
\end{center}
\caption{\label{fig:data_roi_color} 
Maps of the observed flux by the Fermi-LAT in the energy range $1-100$ GeV, in units of photons cm$^{-2}$ s$^{-1}$, for the four DM profiles studied. 
Upper left: Einasto, upper right: NFW, bottom left: NFW$_c$, and bottom right: Burkert.
For each profile, the ROI is the region inside the circle excluding the band on the Galactic plane.
Color scale is logarithmic, yellow, red and blue correspond to $3.6\times10^{-9}$, $6.4\times10^{-10}$ and $1.2\times10^{-10}$ 
photons cm$^{-2}$ s$^{-1}$, respectively. These values also correspond to black contours. Reprinted from Ref.~\citen{Gomez-Vargas:2013bea}.}
\end{figure*}

\begin{figure*}[!ht]
\begin{center}
\includegraphics[width=0.45\textwidth]{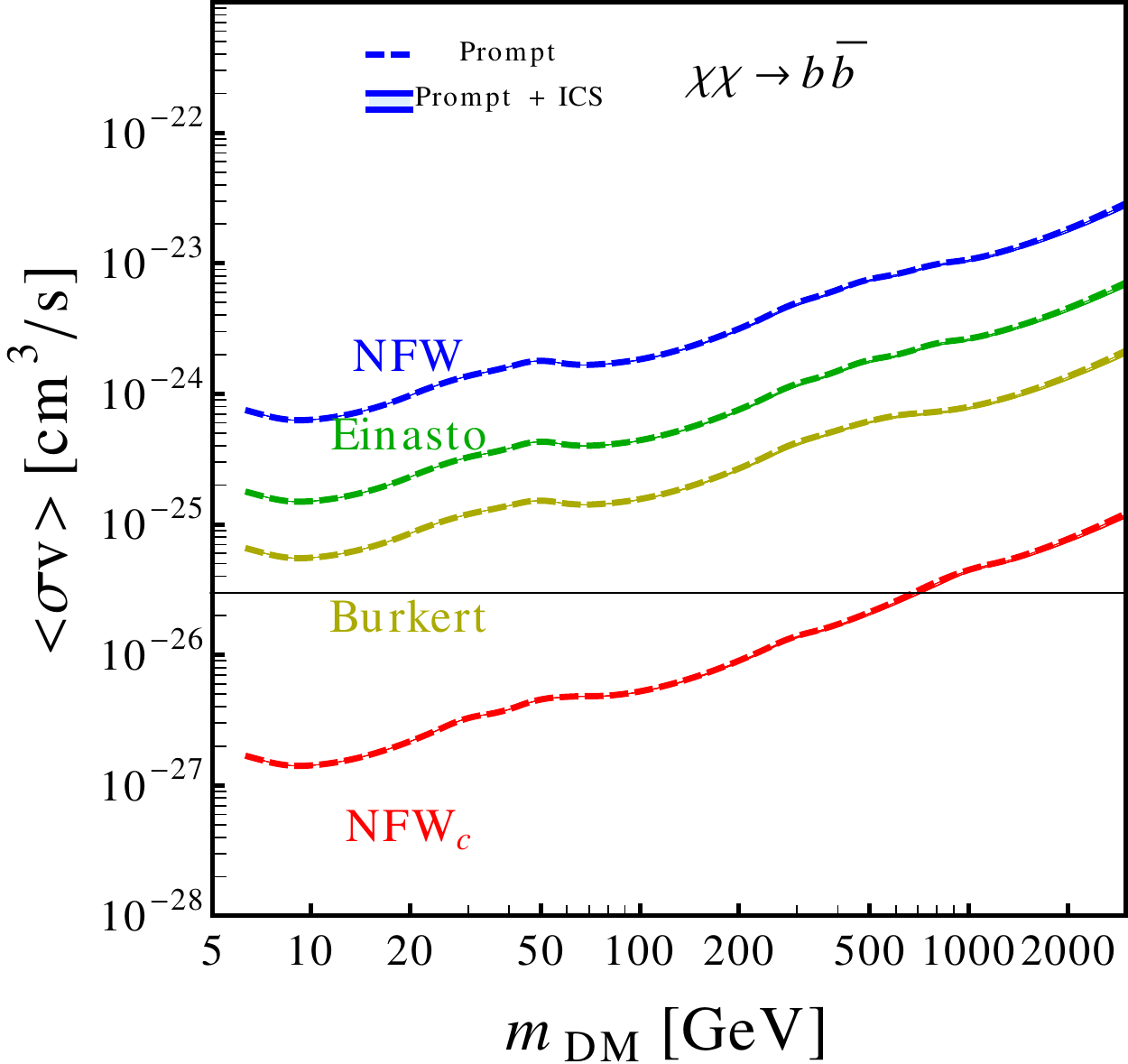}
\includegraphics[width=0.45\textwidth]{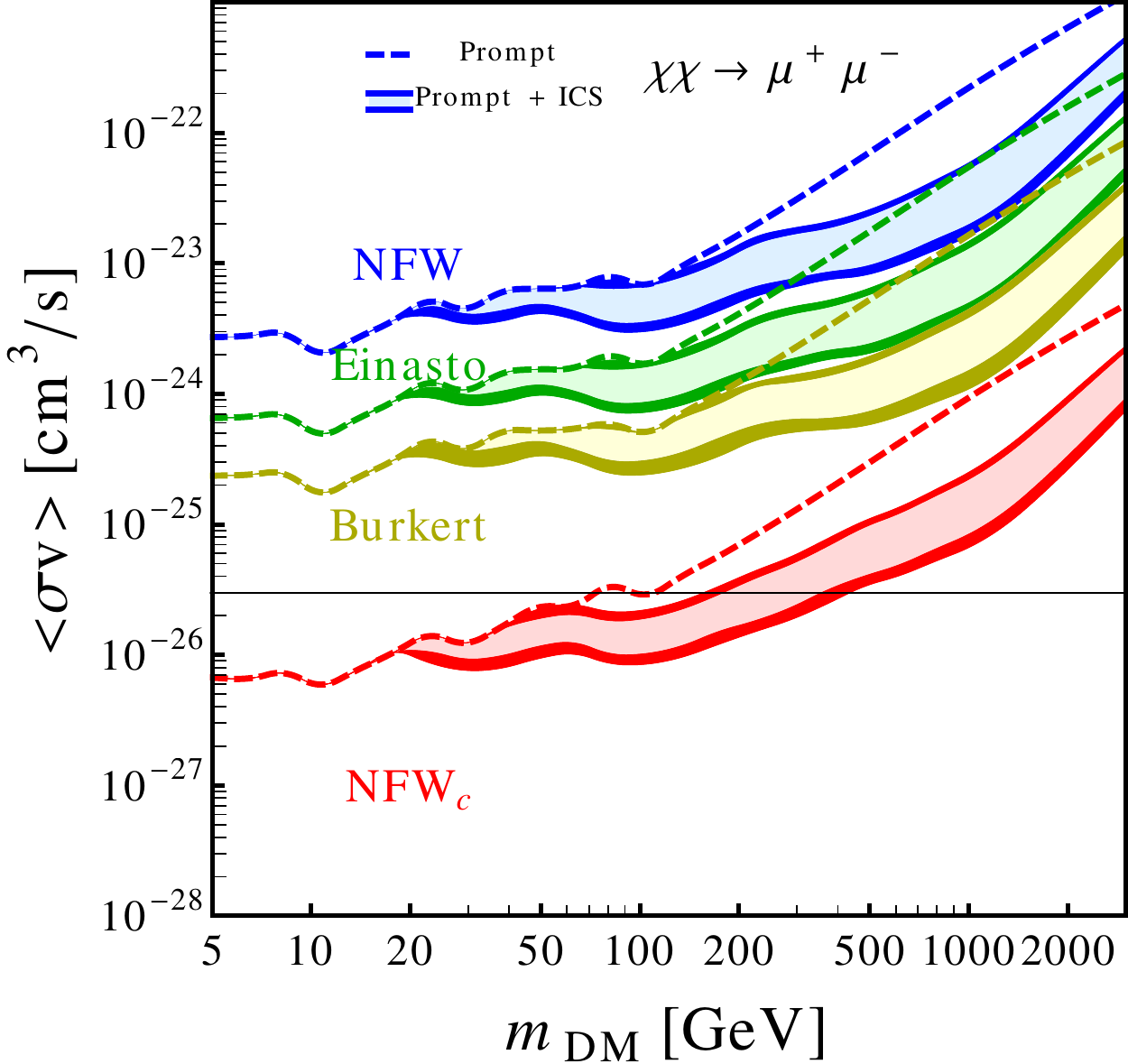}
\includegraphics[width=0.45\textwidth]{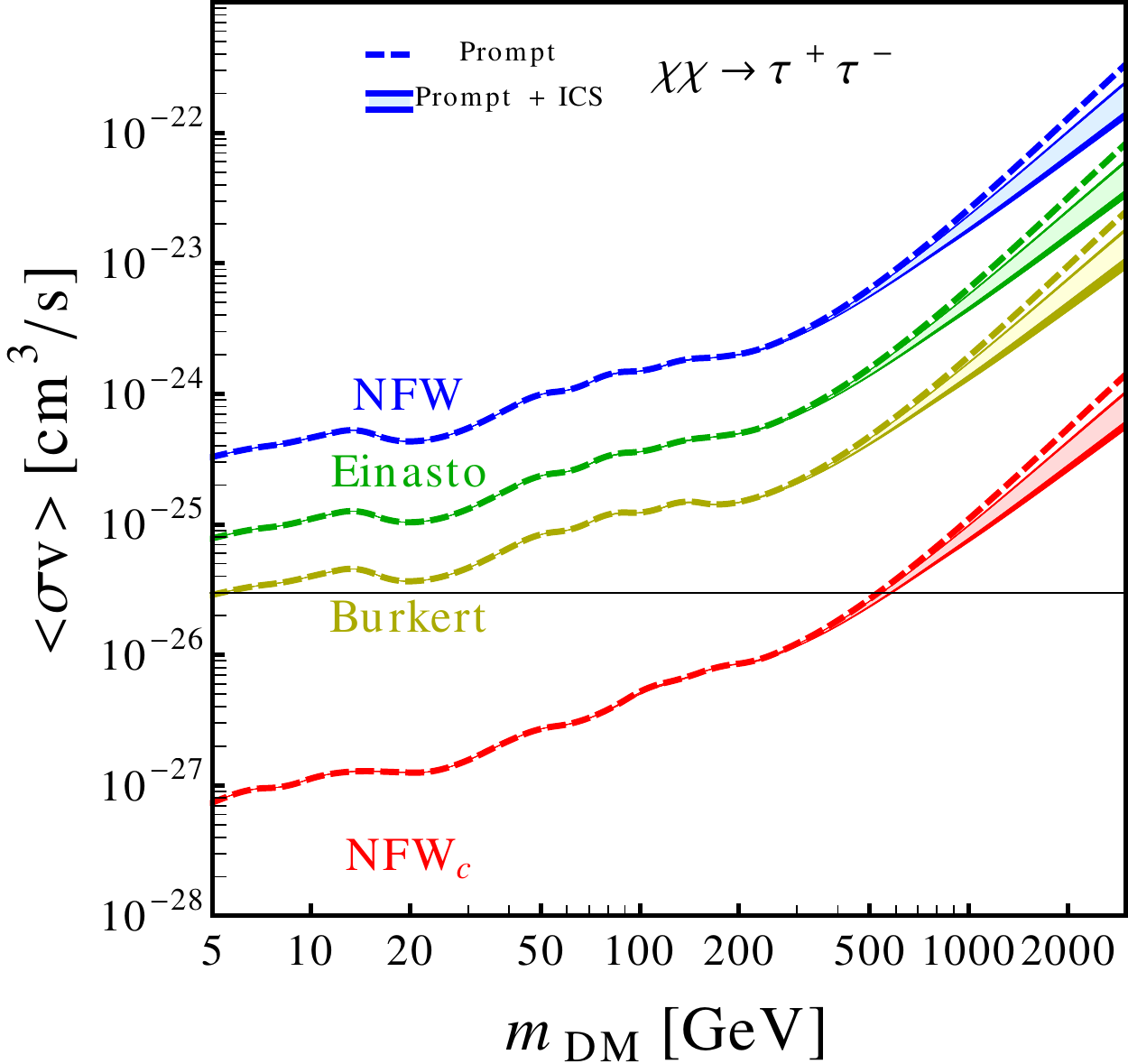}
\includegraphics[width=0.45\textwidth]{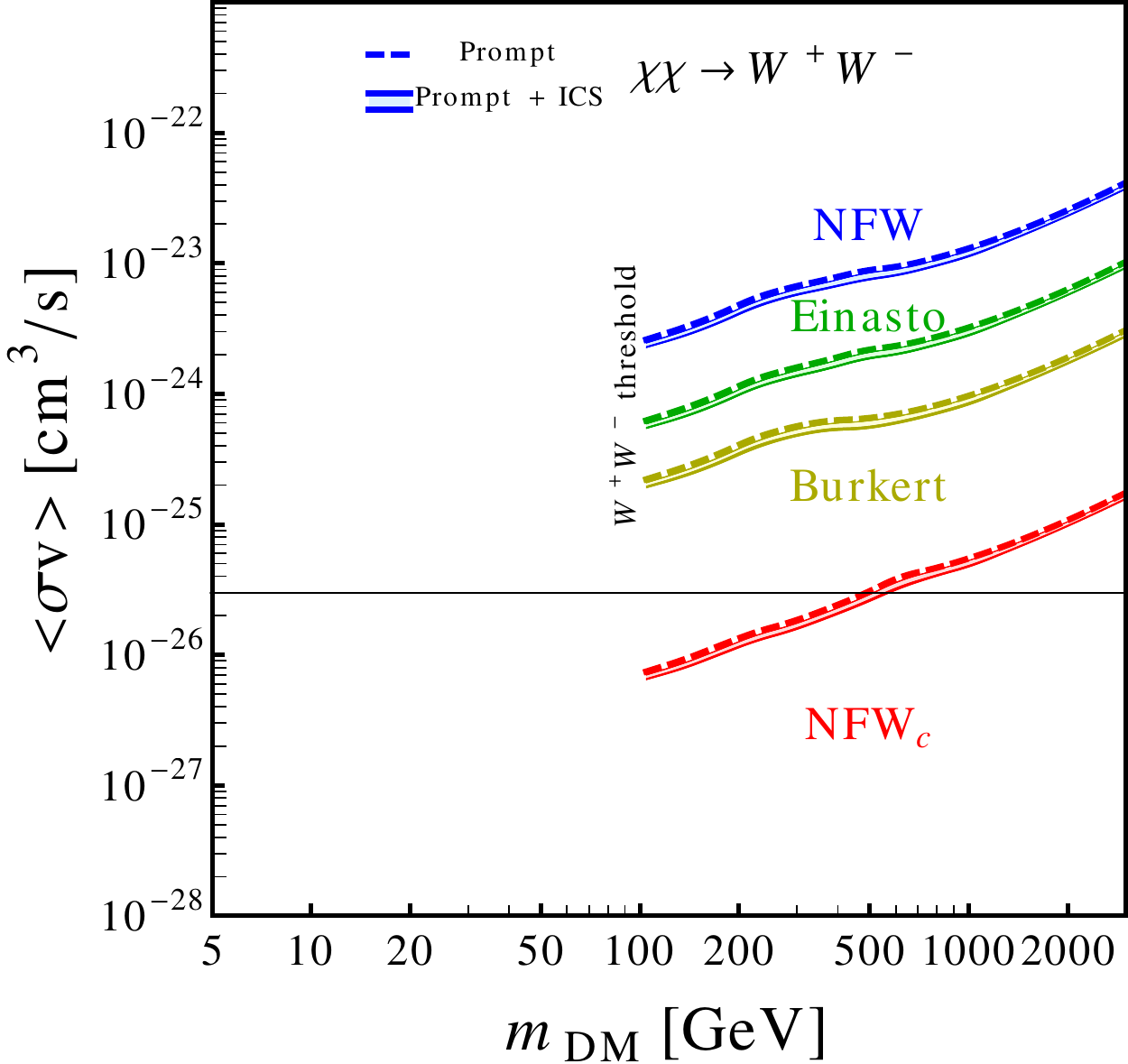}
\end{center}
\caption{\label{fig:const_b} $3\sigma$ upper limits on the annihilation cross-section of models in which DM annihilates into $b\bar b$, $\mu^+\mu^-$ (upper panel), $\tau^+\tau^-$ or $W^+W^-$ (lower panel), for the four DM density profile models. Upper limits set without including the ICS component in the computation are also given as dashed curves (prompt) for comparison. The uncertainty in the diffuse model is shown as the thickness of the solid curves, while the lighter shaded regions represent the impact of the different strengths of the Galactic magnetic field with lower(higher) values of the cross-section corresponding to $B_0=1~\mu$G($B_0=10~\mu$G). The horizontal line corresponds to the expected value of the thermal cross-section for a generic WIMP candidate. Reprinted from Ref.~\citen{Gomez-Vargas:2013bea}.} 
\end{figure*}

\subsection{Constraints from the analysis of the Galactic Center} 
The likely brightest source of gamma rays from DM annihilation is the center of our Galaxy. 
However, the line-of-sight to the GC traverses the galactic disc, which harbours numerous 
highly-energetic processes ($\pi_0$ production in cosmic-ray interactions, bremsstrahlung 
and inverse Compton emission, bright point sources). Furthermore, the uncertainties in 
the signal and background morphologies make the identification of a DM signal from the 
inner Galaxy a challenging task.

A conservative limit on DM can be set assuming that all gamma-ray emission in a region around 
the GC might come from dark matter. Therefore the expected DM signal must not exceed the observed 
gamma-ray emission\cite{Gomez-Vargas:2013bea}. The target regions in 
Ref.~\citen{Gomez-Vargas:2013bea} were optimized according to four dark matter profile 
models, i.e. the Navarro-Frenk-White (NFW) profile~\cite{nfw}, the Einasto 
profile \cite{einasto,navarro04}, the Burkert profile \cite{burkert} and 
a compressed NFW (NFWc) profile. Figure \ref{fig:data_roi_color} shows the target 
regions optimized for each DM density profile, overlayd with the observed flux 
measured by the Fermi-LAT in the energy range $1-100$ GeV.

Figure~\ref{fig:const_b} shows the constraints obtained for different final states. The 
constraints obtained in the likely case that the collapse of baryons to the Galactic Center 
is accompanied by the contraction of the dark matter are strong. In particular, 
for the $b\bar b$ and $\tau^+\tau^-$ or $W^+W^-$ dark matter annihilation channels, the 
upper limits on the annihilation cross section imply that the thermal cross section is 
excluded for a WIMP mass smaller than about 700 and 500 GeV respectively. For 
the $\mu^+ \mu^-$ channel, where the effect of the inverse Compton scattering is 
important, depending on the models of the Galactic magnetic field, the thermal cross-section 
is excluded for a WIMP mass smaller than about 150 to 400 GeV.

The upper limits on the annihilation cross section of dark matter particles obtained 
are two orders of magnitude stronger than those without contraction. Improved modeling 
of the Galactic diffuse emission as well as the potential contribution from other astrophysical 
sources (for instance unresolved point sources) could provide more constraining limits.

\begin{figure*}[!b] 
\begin{center}
\includegraphics[width=0.48\textwidth]{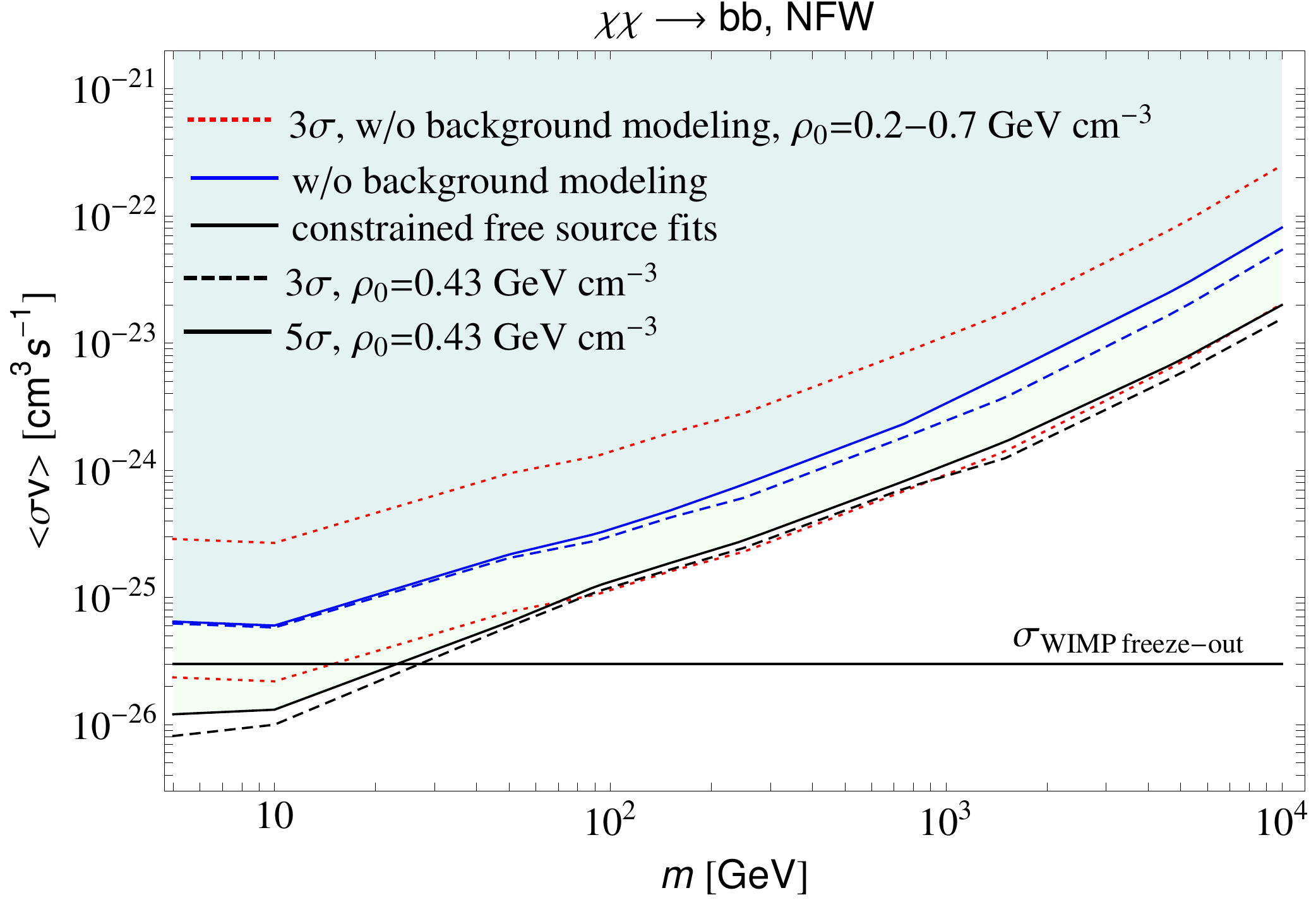}
\includegraphics[width=0.48\textwidth]{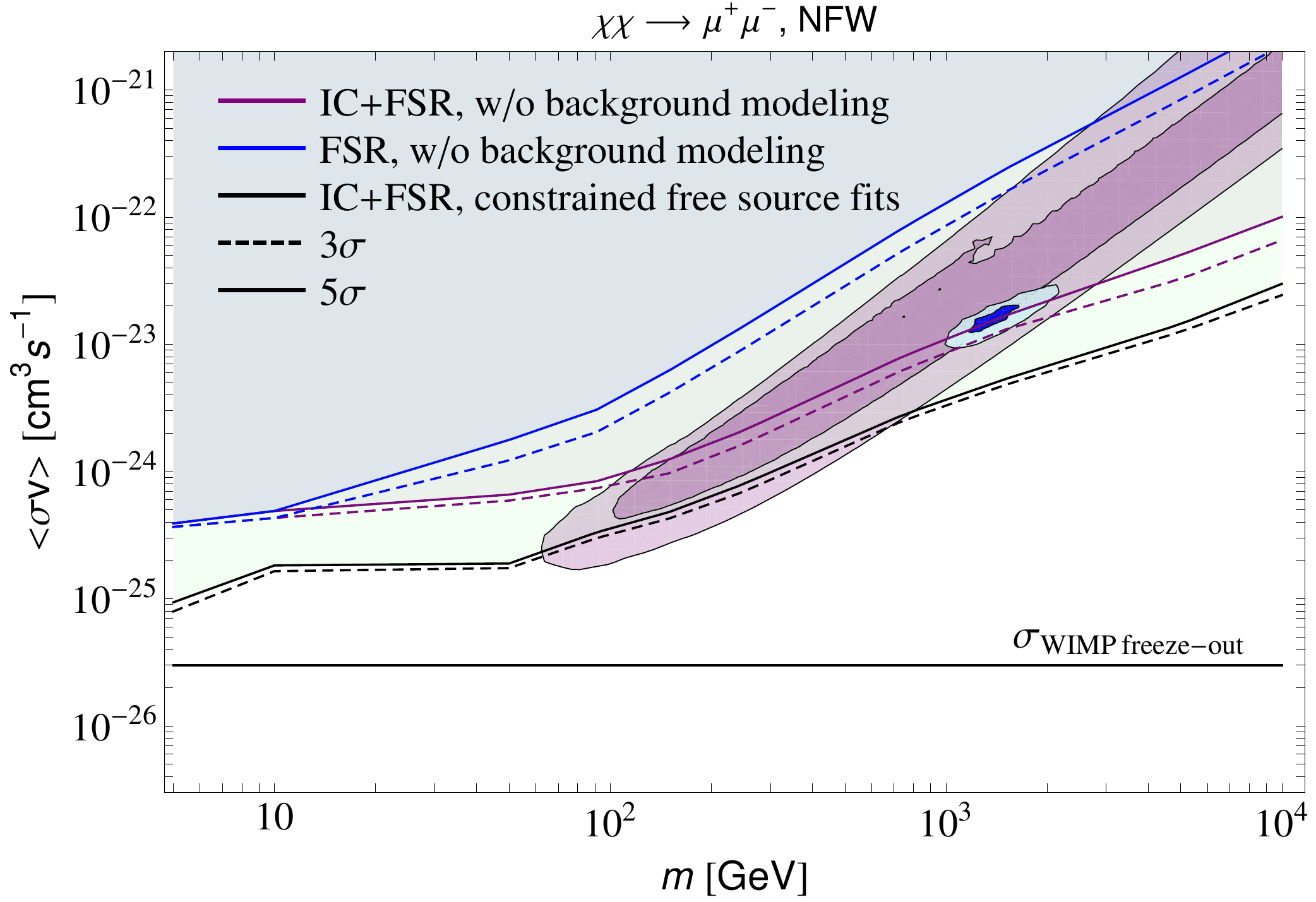}
\includegraphics[width=0.5\textwidth]{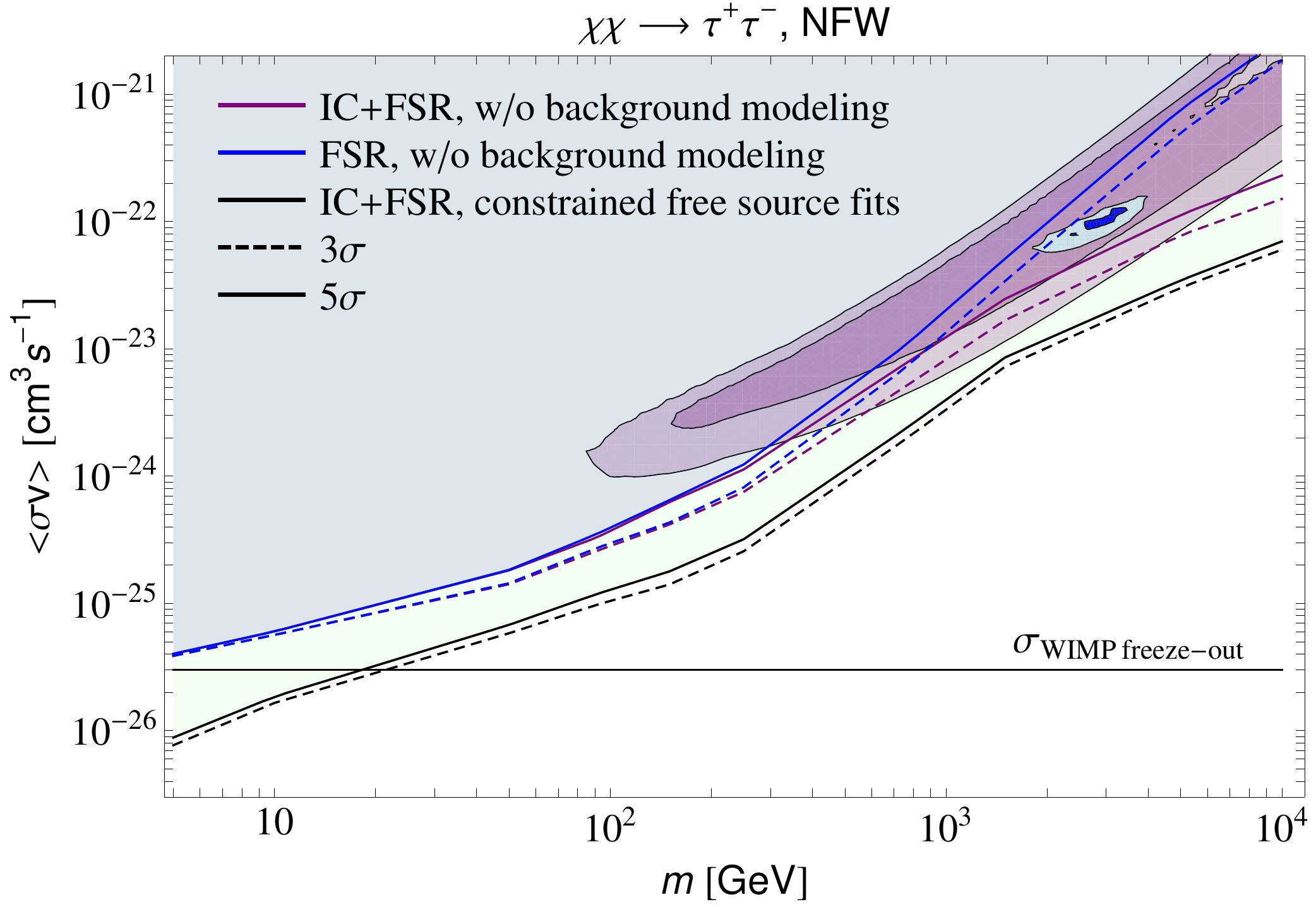}
\end{center}
\caption{Upper limits on the velocity averaged DM annihilation cross-section including a model of the astrophysical background compared with the limits obtained with no modeling of the background. Upper panel: Limits on models in which DM annihilates into $b{\bar b}$ (left) and to $\mu ^+ \mu^-$ (right), for a DM distribution given by the NFW distribution. In the left panel it is also shown an uncertainty band (red dotted lines) in the $3\sigma$ no-background limits  which would result from varying the local DM density $\rho_0$ in the range 0.2-0.7 GeV cm$^{-3}$. Lower plot: The same, for DM annihilation to $\tau ^+ \tau^-$. The horizontal line marks the thermal decoupling cross section expected for a generic WIMP candidate. The region excluded by the analysis with no model of the astrophysical background is indicated in light blue, while  the additional region excluded by the analysis with a modeling of the background is indicated in light green. The regions of parameter space which provide a good fit to PAMELA \cite{Adriani:2008zr,Adriani:2008zq} (purple) and Fermi-LAT \cite{Abdo:2009zk,Ackermann:2010ij} (blue) CR electron and positron data are shown, as derived in \cite{2010NuPhB.840..284C} and are scaled by a factor of 0.5, to account for different assumptions on the local DM density. Reprinted from Ref.~\citen{latdmhalo}.}
\label{fig:fixedsourcelimits}
\end{figure*}

\subsection{Constraints from the analysis of the Galactic halo} 
In order to minimize uncertainties connected with the region of the Galactic Center, 
the analysis in Ref.~\citen{latdmhalo} considered a region of interest consisting of two 
off-plane rectangles ($5^\circ \leq |b| \leq 15^\circ$ and $|l| \leq 80^\circ$) and searched 
for a continuous emission from DM annihilation or decay in the smooth Galactic dark matter 
halo. They implemented two different approaches. In the first case a conservative 
choice was taken and limits were set on DM models assuming that all the gamma ray emission 
in the region might come from dark matter (i.e. no astrophysical signal is modeled and subtracted). 
In the second approach dark matter source and astrophysical emission were fit simultaneously 
to the data, marginalizing over several relevant parameters of the astrophysical emission. 
As no robust signal of DM emission was found, DM limits were set (see Fig.~\ref{fig:fixedsourcelimits}).

These limits are particularly strong on leptonic DM channels, which are hard to constrain 
in most other probes (notably in the analysis of the dwarf Galaxies, described below). 
This analysis strongly challenges DM interpretation\cite{grasso} of the positron rise
observed by PAMELA~\cite{Adriani:2008zr,Adriani:2008zq} and the $e^{\pm}$ spectrum observed by Fermi LAT~\cite{Abdo:2009zk,Ackermann:2010ij}.

\begin{figure*}[!ht]
\centering
\includegraphics[width=0.9\textwidth]{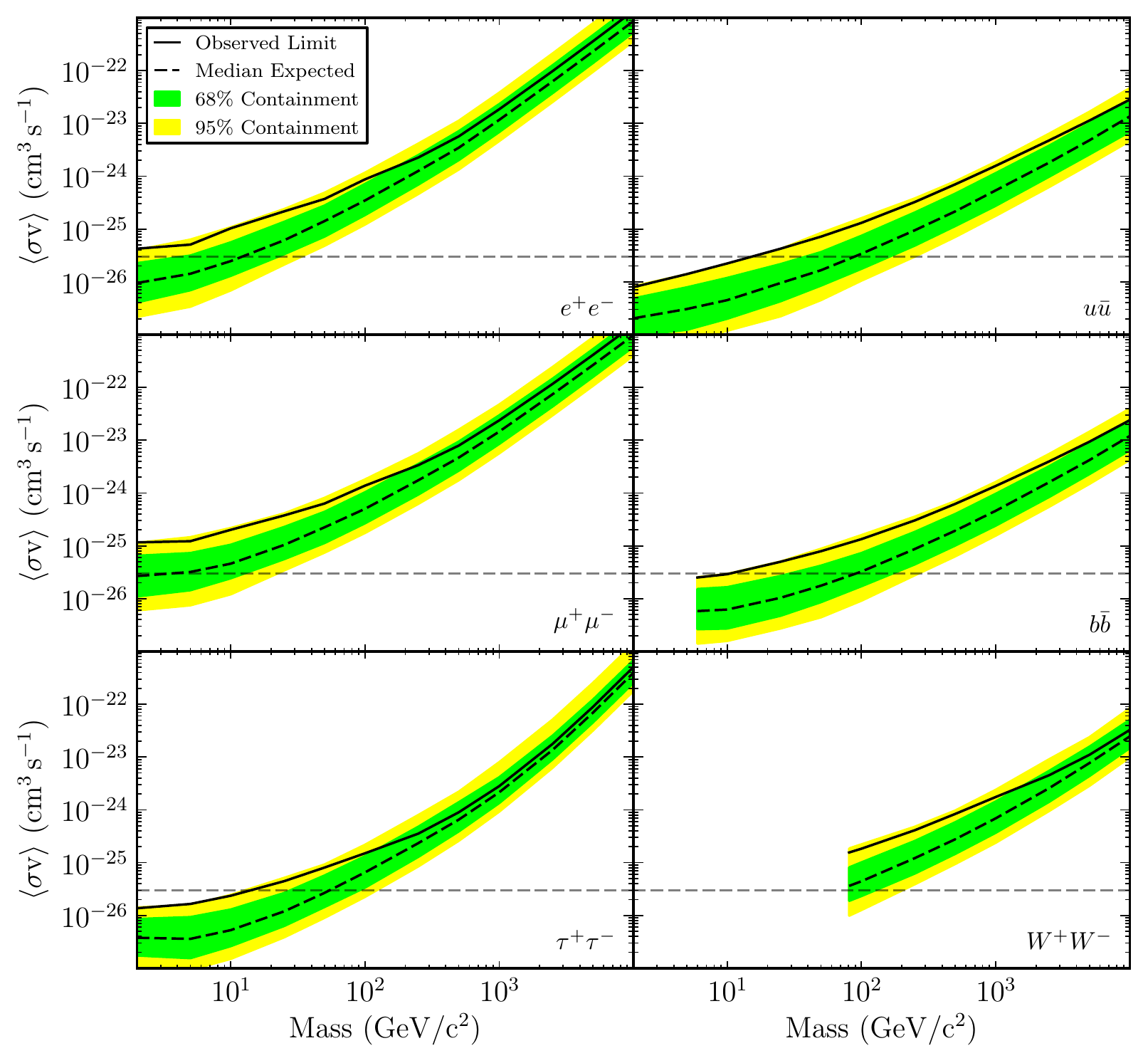}
\caption{Constraints on the dark matter annihilation cross section at 95\% CL derived from a combined analysis of 15 dwarf spheroidal galaxies assuming an NFW dark matter distribution (solid line). 
In each panel bands represent the expected sensitivity as calculated by repeating the combined analysis on randomly-selected sets of blank fields at high Galactic latitudes in the LAT data.
The dashed line shows the median expected sensitivity while the bands represent the 68\% and 95\% quantiles. The positions and widths of the expected sensitivity bands reflect the range of statistical fluctuations expected both from the LAT data and from the stellar kinematics of the dwarf galaxies. Reprinted from Ref.~\citen{Ackermann:2013yva}.}
\label{fig:15dwarf}
\end{figure*}

\subsection{Constraints from the analysis of dwarf spheroidal galaxies} 
Many complications associated with the GC are avoided when looking at point-like targets 
outside the galactic disk. The corresponding signals are typically considerably fainter, 
which is however compensated by the greatly simplified and much smaller astrophysical background. 
Probably most promising source candidate target class are nearby Milky Way dwarf spheroidal 
(dSph) galaxies. This is because their mass-to-light ratio is predicted to be from 10 
to 1000~\cite{wolf, simon}, implying that they could be largely DM dominated. 
Moreover, since no significant gamma-ray emission of astrophysical origin is expected 
(these systems host few stars and no hot gas), the detection of a gamma-ray signal could 
provide a clean DM signature. 

The LAT detected no significant emission from any of such systems and the upper limits 
on the $\gamma$-ray flux allowed us to put very stringent constraints on the parameter 
space of some well motivated WIMP models. In particular, a combined analysis of all 
known dwarf satellites with the Fermi-LAT, that also incorporates the uncertainties 
in the J-factors for these objects, has pushed the annihilation cross section limits 
below the canonical thermal relic production cross-section for a range of WIMP masses 
(around 10 GeV) for the annihilation into some channels which often act as a 
benchmark~\cite{lat_DMdwarf_paper, geringer_DMdwarf, Mazziotta:2012ux,Ackermann:2013yva}. 

\begin{figure}[!ht]
\includegraphics[width=0.9\textwidth]{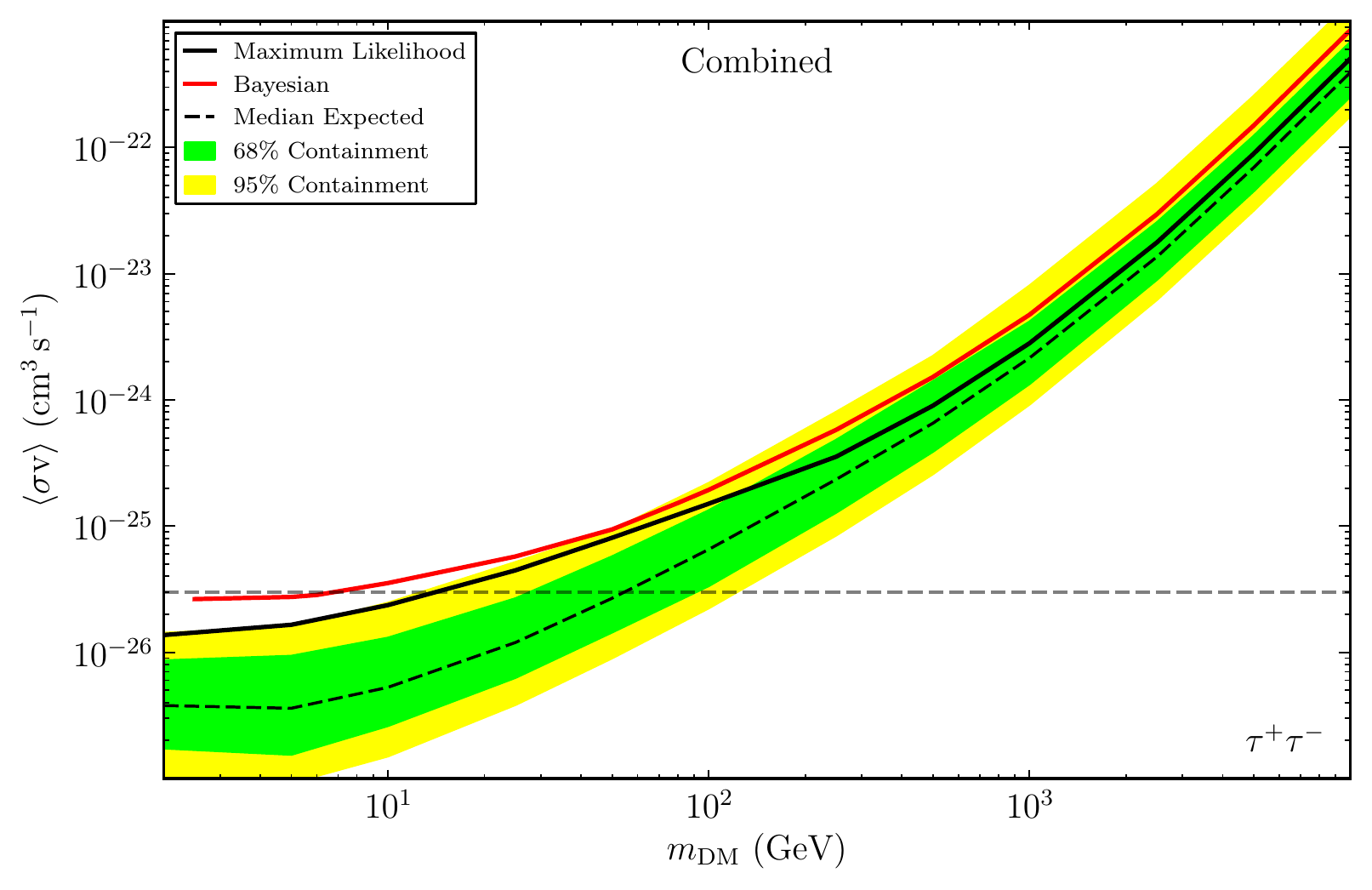}
\caption{Comparison of constraints on the dark matter annihilation cross section ($\tau ^+ \tau^-$ channel) derived from the combined maximum likelihood and the combined Bayesian analyses of 15 dwarf spheroidal galaxies. The expected sensitivity for the maximum likelihood analysis is represented similarly to Fig.~\ref{fig:15dwarf} The observed Bayesian limits are consistent with the expected Bayesian sensitivity bands (not shown), which are likewise higher than those of the maximum likelihood analysis. Reprinted from Ref.~\citen{Ackermann:2013yva}.}
\label{fig:15dwarfbaye}
\end{figure}

The main advantages of the combined analysis approach are that the analyses can be individually 
optimized and that combined limits are more robust under individual background fluctuations 
and under individual astrophysical modelling uncertainties than individual limits. 
Moreover, different analysis methods are also employed to set robust results.

A combined likelihood analysis of the 15 most promising dwarf galaxies, based on 4 years
of data and pushing the limits below the thermal WIMP cross section for low DM
masses (below a few tens of GeV), has been recently performed in Ref.~\citen{Ackermann:2013yva} 
(see Fig.~\ref{fig:15dwarf}). A novel technique is used to obtain J-factors for the 
dwarf spheroidal galaxies by deriving prior probabilities for the dark matter distribution 
from the population of Local Group dwarf galaxies\cite{Martinez:2013els}. A more 
advanced statistical framework to examine the expected sensitivity of thes search was devolped 
in Ref.~\citen{Ackermann:2013yva}. An extensive study of systematic effects arising 
from uncertainties in the instrument performance, diffuse background modeling, 
and dark matter distribution was also performed. A set of random blank-sky locations 
as a control sample for the analysis of dwarf spheroidal galaxies was used, introduced for 
the first time by Ref.~\citen{Mazziotta:2012ux} to analyze the Milky Way dark matter halo.

As a final check, a Bayesian analysis based on the work of Ref.~\citen{Mazziotta:2012ux} was 
also included in Ref.~\citen{Ackermann:2013yva}, by folding the dark matter signal spectrum 
with the LAT instrument response function and incorporating information from all energy bins 
when reconstructing a posterior probability distribution for the dark matter cross section. 
This approach derives the diffuse background empirically from annuli surrounding the dwarf galaxies. 
Despite differences in background modeling, the treatment of the LAT instrument performance, 
and the methodology for setting upper limits, the Bayesian and maximum likelihood analyses 
provide comparable results (see Fig.~\ref{fig:15dwarfbaye}). The discovery and characterization 
of new dwarf spheroidal galaxies could greatly improve the LAT sensitivity to dark matter annihilation 
in this class of objects, in particular at high DM mass values.

\begin{figure}[!ht]
\begin{center}
\includegraphics[width=0.9\textwidth]{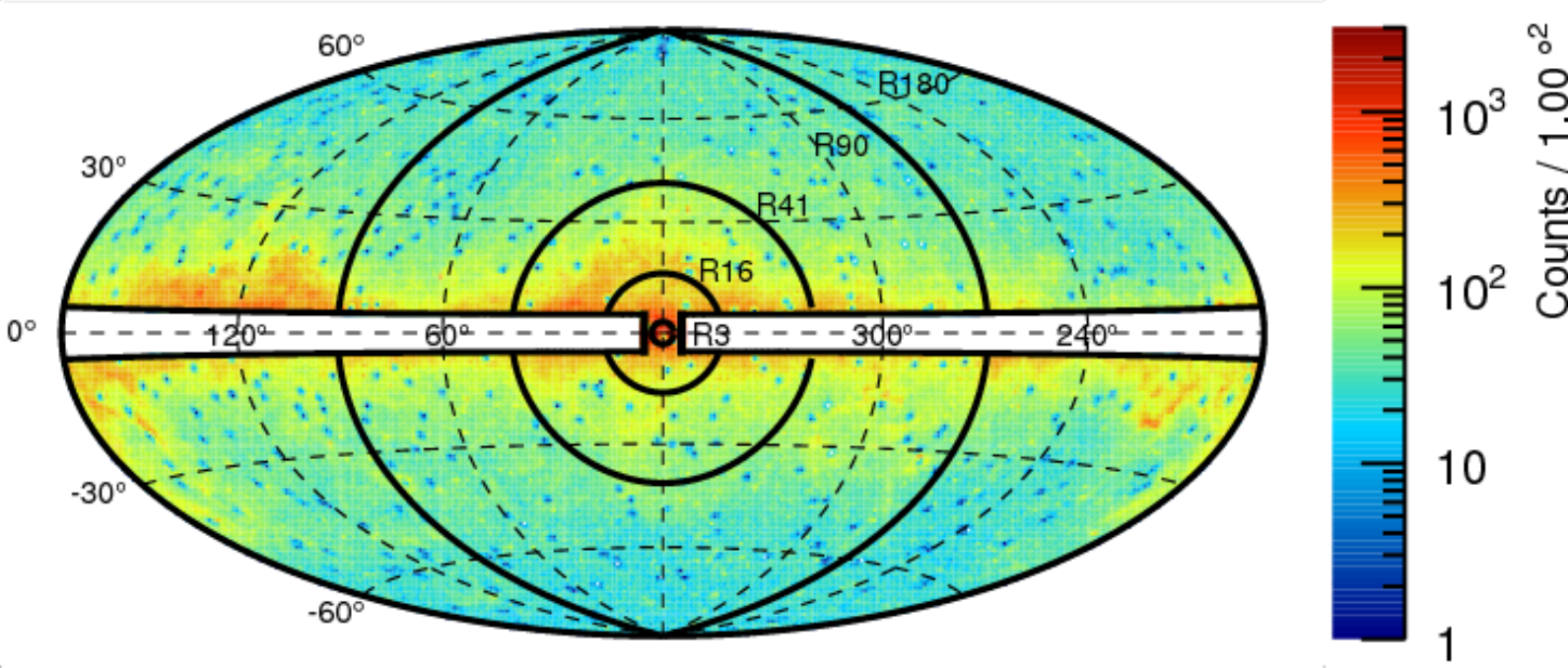}
\end{center}
\caption{Counts map for the line search dataset
  binned in $1^{\circ} \times 1^{\circ}$ spatial bins in the R180
  ROI, and plotted in Galactic coordinates using the Hammer-Aitoff
  projection.  The energy range is 2.6--541~GeV and the most-significant
  2FGL sources have been removed using an energy-dependent
  mask.  Also shown are the outlines of
  the other ROIs (R3, R16, R41, and R90) used in Ref.~\citen{Fermi-LAT:2013uma}. Reprinted from Ref.~\citen{Fermi-LAT:2013uma}.}
\label{fig:linerois}
\end{figure}

\subsection{Gamma-ray lines}
A line at the WIMP mass, due to for instance to the 2$\gamma$ production channel, 
could be observed as a feature in the astrophysical source spectrum\cite{balt}. 
Such an observation in the high energy region would be a “smoking gun” for WIMP DM 
as it is difficult to explain by a standard astrophysical origin of $\gamma$ emission.
If a WIMP $\chi$ annihilates or decays directly into a photon $\gamma$ and another 
particle $X$, the photons are approximately monochromatic with energy 
\begin{equation}
E_{\gamma}=m_{\chi}\,\Big(1-\frac{m^2_X}{4m_{\chi}^2}\Big)
\end{equation}
for annihilations and replacing $m_{\chi}\rightarrow m_{\chi}/2$ for decays 
(we assume that the typical velocity of $\chi$ is very small, of the order of $v/c\sim 10^{-3}$, 
therefore these signals should be approximately monochromatic in the lab frame as well). 
Additionally, gamma rays created in WIMP annihilations via internal bremsstrahlung could 
produce a sharp spectral feature near the mass value of the $\chi$~\cite{Bringmann:2007nk}.

No significant evidence of gamma-ray line(s) was found using 11 months and 2 years 
of LAT data~\cite{Abdo:2010nc,Ackermann:2012qk}. Recently, the claim of an indication of 
a line-like feature at 130~GeV in the Fermi-LAT data has drawn considerable attention. 
This feature is reported to be strongly correlated with the Galactic center 
region~\cite{Bringmann:2012vr,Weniger:2012tx,Tempel:2012ey,Su:2012ft}, and also 
with nearby galaxy clusters~\cite{Hektor:2012kc}, and unassociated LAT 
sources~\cite{Su:2012zg,Hektor:2012jc}. The feature has not been seen in the vicinity of 
nearby dwarf galaxies~\cite{GeringerSameth:2012sr}. However such a signal is expected 
to be much fainter than in the Galactic center. A systematic investigation of the spatial 
morphology of the 130~GeV feature and other line-like features in the Galactic plane 
is presented in~\cite{Boyarsky:2012ca}.

Potential instrumental effects and a similar feature detected in the bright $\gamma$-ray 
emission from cosmic-ray (CR) interactions in Earth's upper atmosphere (the Limb) have also 
been discussed~\cite{Whiteson:2012hr,Hektor:2012ev,Finkbeiner:2012ez}. 

\begin{figure}[!ht]
\begin{center}
\includegraphics[width=0.8\textwidth]{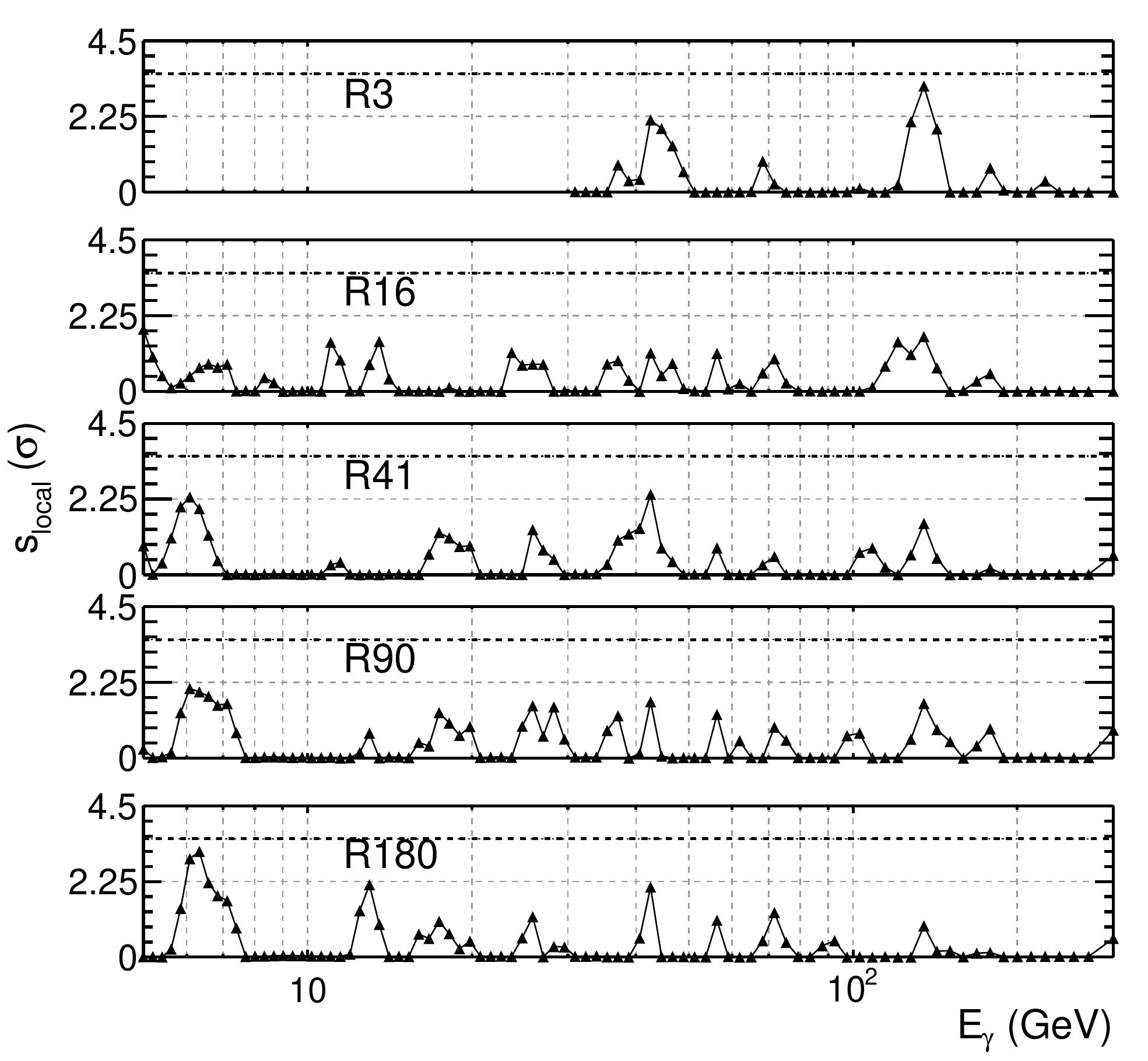}
\end{center}
\caption{Local fit significance vs line energy in all five ROIs.
Note that line signal was required to be non-negative. The dashed line
at the top of the plot indicates the local significance corresponding
to the 2 $\sigma$ global significance taking into account the trial effect. Reprinted from Ref.~\citen{Fermi-LAT:2013uma}.}
\label{fig:summarylines}
\end{figure}

In the recent analysis of the 4 year data~\cite{Fermi-LAT:2013uma} the Fermi LAT team 
has improved over the two-year paper in three important aspects:
\begin{itemize}
\item a new improved data set (Pass-7 reprocessed) was used, as it corrects for 
losses in the calorimeter light yield due to radiation damage during the four years of 
the Fermi mission
\item regions of interest (ROIs) were selected \textit{a priori} to maximize the 
sensitivity based on different DM density profiles
\item an event-by-event estimate of the energy reconstruction quality in the parametrization 
of the energy dispersion was also included (so called 2D PDF fitting method in Ref.~\citen{Fermi-LAT:2013uma}).
\end{itemize}

A set of five ROIs were optimized in Ref.~\citen{Fermi-LAT:2013uma} for sensitivity to WIMP
annihilation or decay and four reference models for the distribution of DM in the Galaxy were
considered. The ROIs were defined as a circular regions of radius $R_{gc}$
centered on the Galactic center with a mask corresponding to the region $|b| < 5^{\circ}$ 
and $|l| > 6^{\circ}$, which were optimized for each of the DM density profiles considered. 
For annihilating DM models they used $R_{gc} = 3^{\circ}$ (R3), optimized 
for the contracted NFW profile, $R_{gc} = 16^{\circ}$ (R16) optimized for 
the Einasto profile, $R_{gc} = 41^{\circ}$ (R41), optimized for the NFW and
$90^{\circ}$ (R90), optimized for the Isothermal profile. 
For decaying DM models they used all the sky, i.e. $R_{gc} =180^{\circ}$ 
(R180) (see Fig.~\ref{fig:linerois}).

\begin{figure}[!ht]
\begin{center}
\includegraphics[width=0.9\textwidth]{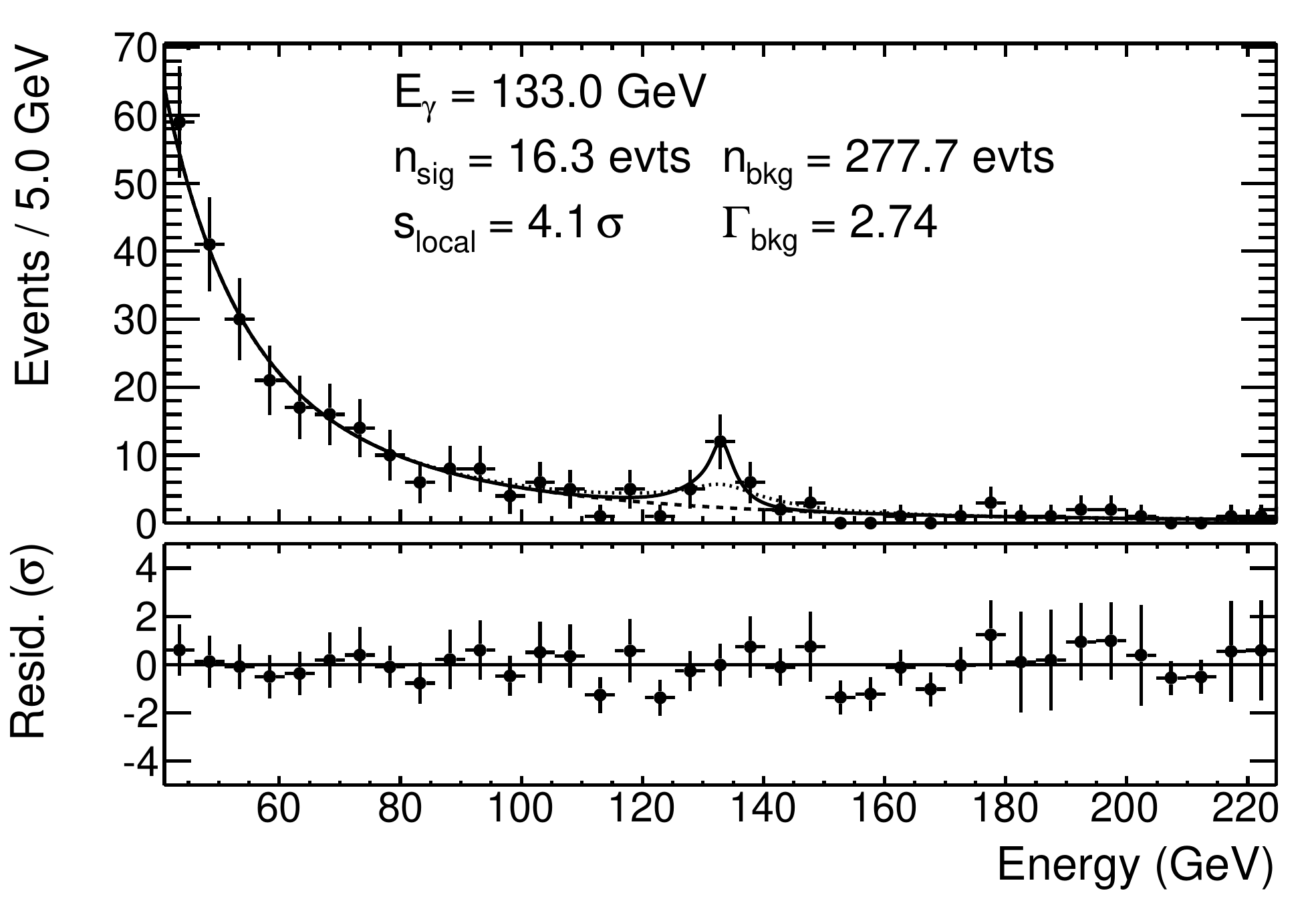}
\end{center}
\caption{Fit to a $\gamma$-ray line at 133 GeV in the R3 using the 2D PDF fitting method, including a scale factor for the width of the energy dispersion. The solid curve shows the average model of the fitted events. The best-fit width of the energy resolution was $0.32^{+0.11}_{-0.07}$(95\% C.L.). The dotted line shows the best-fit curve with the energy dispersion width fixed to 1.0. Note that these fits were unbinned; the binning here is for visualization purposes. Reprinted from Ref.~\citen{Fermi-LAT:2013uma}.}
\label{fig:linewidth}
\end{figure}

In that analysis no globally significant lines have been found and new limits 
on the $\gamma-\gamma$ DM annihilation channel were set (see Fig.~\ref{fig:summarylines}). 
In a close inspection the 130 GeV feature, it was found that indeed there is a feature 
around 133 GeV (due to the new LAT calibration constants), with a signal in R3 at 
about 4 $\sigma$ local significance by using the the ’1D’ PDF fitting method 
(i.e. without including the quality of the energy of the energy dispersion). However, 
the significance drops to 3.3 $\sigma$ local, or $\leq 2$ sigma global significance 
once trials factors are taken into account, by using the 2D PDF fitting method. 
In addition, the feature is too narrow with respect to the energy dispersion of the LAT instrument. 
In fact, to quantify this, they scaled the width of the energy dispersion by a scale factor, 
then they refitted the feauture at 133 GeV the best-fit value of this scaling 
factor was $0.32^{+0.11}_{-0.07}$(95\% C.L.), as shown in Fig.~\ref{fig:linewidth}.

In addition, a weaker signal is found at the same energy in the control sample 
(in the Earth limb), which might point to a systematic effect present in this data set. 
In order to examine this possibility weekly observations of the Limb were scheduled, 
and a better understanding of the nature of the excess in the control sample is expected.

\section{Indirect Dark Matter searches with electrons and positrons}

If the standard model particles resulting from DM annihilation or decays are charged CRs, 
they do not travel directly to us. Instead, they are transported to the Solar System 
via scattering on $\mu G$ galactic magnetic field (GMF) irregularities in the interstellar 
medium (ISM) and halo surrounding the Galaxy. Their trajectories are quickly randomized 
by such processes so that they retain little information about their initial directions. 
This happens because the Larmor radius for a typical value of 4 $\mu G$ for 
the GMF and for a 100 GeV singly-charged particle is $\sim 3 \times 10^{-5}$ pc, 
considerably smaller than the typical distance to a nearby source (of the order of a hundred pc).

For energies $>$ 10 GeV, the energy losses of the CR nuclei are strongly suppressed compared 
to the lighter electrons and positrons. Hence, the main effect on the CR nuclei is from 
scattering. Particles produced throughout the halo, at distances of tens of kiloparsecs 
and further, can reach the Solar System. Electrons and positrons, however, are severely affected 
by IC scattering on the interstellar radiation field (ISRF) and by synchrotron radiation 
from spiraling in the Galactic magnetic field. If produced with energies $>$ 100 GeV, 
they will reach the Solar System only if their origin is within a few kiloparsecs. 

Cosmic-ray electrons/positrons (CREs) with energies 100 GeV (1 TeV) can be observed at 
the Earth if the are originated from relatively nearby locations, less than about 1.6 kpc 
(0.75 kpc) away~\cite{Ackermann:2010ip}. This means that it could be possible that such 
high energy CREs originate from a highly anisotropic collection of a few nearby sources.
As pointed out in the many analyses (see for example Ref.~\citen{grasso}), any anisotropy 
in the arrival directions of the detected CREs is a powerful tool to discriminate between 
a dark matter origin and an astrophysical one.

In addition, in the last decades the searches for a DM signal from the Sun were performed 
looking for possible excesses of neutrinos or gamma rays associated with the Sun’s direction. 
However, as it was noted in Ref.~\citen{Schuster:2009fc}, several DM models that have been 
recently been developed to explain various experimental results, also imply an associated 
solar flux of high-energy cosmic-ray electrons and positrons (CREs). On the other hand, 
no known astrophysical mechanisms are expected to generate a significant high-energy 
CRE ( $>$ 100 GeV) excess associated with the Sun.

\subsection{Anisotropies}
The Fermi-LAT experiment reported high precision measurements of the spectrum of cosmic-ray 
electrons-plus-positrons between 20 GeV and 1 TeV~\cite{Abdo:2009zk,Ackermann:2010ij}. 
The spectrum shows no prominent features, even though it is significantly harder than that 
inferred from several previous experiments. On the other hand, the ATIC experiment 
observed a pronounced bump in the electron + positron spectrum at around 600 GeV~\cite{atic}.
The positron fraction $e^+/(e^+ + e^-)$ has been found to increase with energy above 
10 GeV, first by the PAMELA experiment~\cite{Adriani:2008zq}, then it was 
confirmed by the LAT experiment~\cite{FermiLAT:2011ab} and it was finally measured by 
AMS-02 with high precision~\cite{ams02}.

Early analyses discussed several interpretations of the Fermi results based either 
on a single large scale Galactic CRE component or by invoking additional electron-positron 
primary sources, e.g. nearby pulsars or particle dark matter annihilation~\cite{grasso}. 
Since Galactic dark matter is denser towards the direction of the Galactic center, 
the generic expectation in the dark matter annihilation or decay scenario is a dipole 
with an excess pointing towards the center of the Galaxy and a deficit towards the 
anticenter (see for example Ref.s~\citen{Profumo:2008ms,Hooper:2008kg}).

\begin{figure}[!b]
\begin{center}
\includegraphics[width=0.9\textwidth]{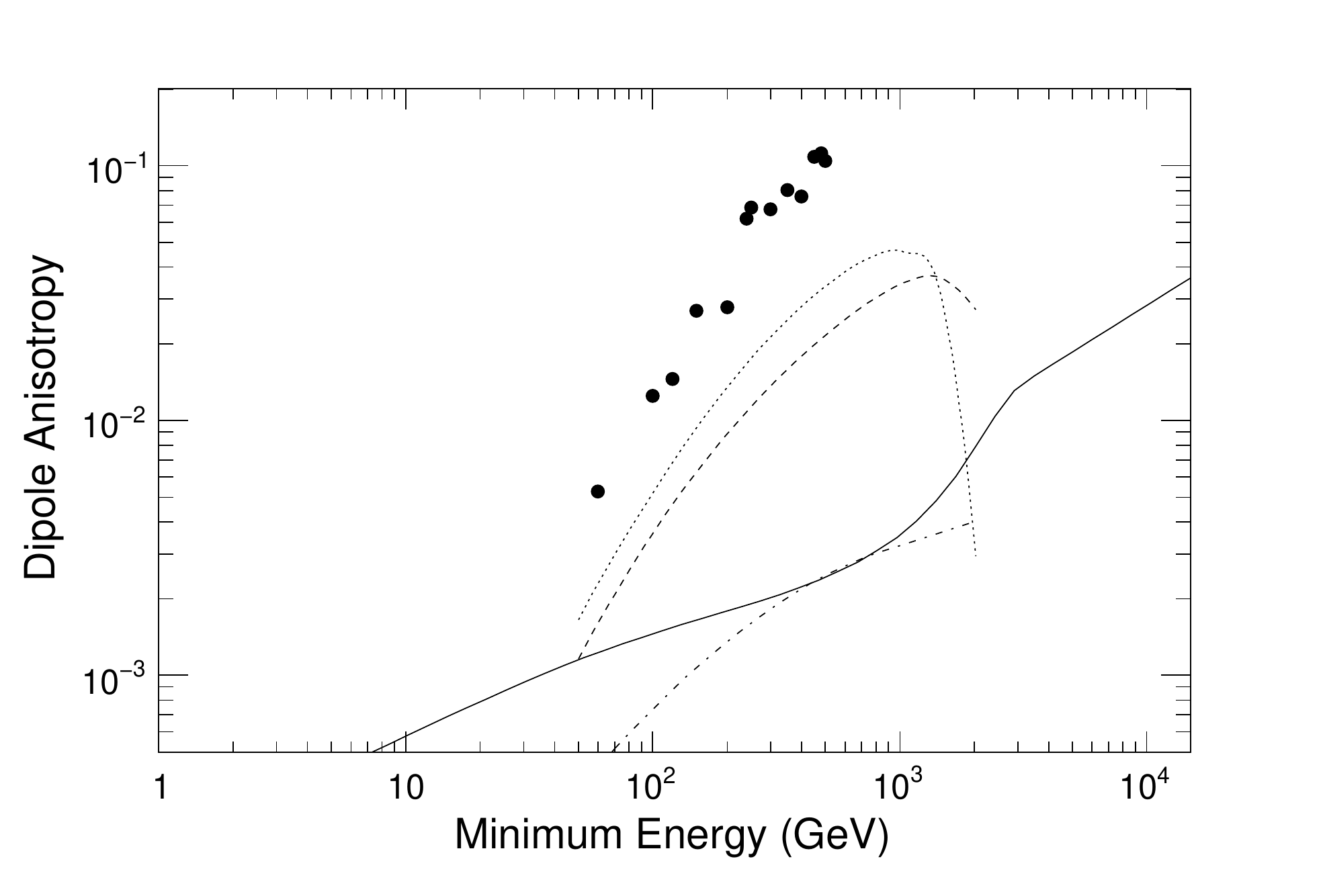}
\end{center}
\caption{Dipole anisotropy versus the minimum energy for some DM scenarios. Solid line: DM distributed in the Milky Way Halo; dashed and  dotted lines: two dark matter benchmark models taken from \cite{Regis:2009qt};  dot-dashed line: DM from the population of Galactic substructures \cite{Cernuda:2009kk}. The 95 \% CL upper limits on the dipole anisotropy evaluated with CRE detected during the first year of LAT instrument are shown with circles. Reprinted from Ref.~\citen{Ackermann:2010ip}.}
\label{fig:anisodm}
\end{figure}

The expected level of dipole anisotropy produced by DM annihilating in the Milky Way halo, 
calculated by tuning the annihilation rate to match the positron fraction measured by the 
PAMELA satellite, is comparable or more likely smaller than the degree of anisotropy 
expected by astrophysical Galactic sources (see Fig.~\ref{fig:anisodm}).
However, there is the possibility that most of the high-energy positrons are produced 
by dark matter annihilations in a nearby dark matter clump. In Ref.~\citen{Regis:2009qt} 
is presented an analysis to evaluate the anisotropy from single nearby dark matter clumps, 
in particular they used two benchmark models that give good fits to the PAMELA and Fermi data. 
In these models, the clumps are moving with a speed of 300 $km~s^{-1}$ perpendicular 
to the Galactic plane, and the dark matter particle has a mass of 5 (3) TeV annihilating 
into $\tau$ leptons, departing at 1.54 kpc or approaching at 1.43 kpc. 
The anisotropy is typically mainly sensitive to the dark matter clumps' distance, 
but the different anisotropies for these two clumps (at almost the same distance) 
are mainly due to an approaching compared to a departing clump (Fig.~\ref{fig:anisodm}). 
Finally, a population of Galactic substructures could produce an anisotropy~\cite{Cernuda:2009kk}.
 
It should be noted that the dark-matter-induced anisotropies predicted here are only valid 
within their given set-up; additional significant CRE-source contributions, a different 
astrophysical background, or a modified diffusion model, would modify the expected 
dipole-anisotropy signal. A detailed study on the dipole anisotropy in the arrival directions 
of high energy CREs due to the Dark Matter can be found in Ref.~\citen{Borriello:2010qh}.

\subsection{Sun}

It is also possible that annihilation of DM trapped within the Sun is a source of 
standard particles. The neutrinos would escape and might be seen in a 
detector~\cite{Silk:1985ax} such as MACRO~\cite{Ambrosio:1998qj}, Super-K~\cite{Desai:2004pq}
or IceCube~\cite{Abbasi:2009uz}. Such a point source with a hard spectrum would be 
convincing evidence that the neutrinos originated from DM. DM builds up in the Sun 
when a WIMP traveling through either body collides with a nucleus and loses enough 
energy to become gravitationally bound. It then orbits the center of the object, 
undergoing multiple collisions. The production rate depends on both the DM annihilation 
cross section and the DM-nucleon scattering cross section. 

A class of models in which DM annihilates to CREs through a new light intermediate
state $\phi$ \cite{Pospelov:2007mp,ArkaniHamed:2008qn} has
been considered to explain the excesses in local CRE fluxes reported
by PAMELA~\cite{Adriani:2008zr}
ATIC~\cite{atic}, and Fermi~\cite{Abdo:2009zk,Ackermann:2010ij}.
In these scenarios DM particles captured by the Sun through elastic scattering interactions
would annihilate to $\phi$ pairs in the Sun's core, and if the $\phi$ could escape
the surface of the Sun before decaying to CREs, these models could
produce an observable CRE flux.

Another class of models in which DM scatters off of nucleons
predominantly via inelastic scattering
has been proposed as a means of reconciling the results of
DAMA and DAMA/LIBRA~\cite{Bernabei:2008yi,Bernabei:2010mq}
with CDMS-II~\cite{Ahmed:2009zw,Ahmed:2010hw} and
other experiments (e.g., Ref.~\citen{Chang:2008gd} and Ref.~\citen{Finkbeiner:2009ug}; see
also Ref.~\citen{Savage:2008er} for a comprehensive discussion of experimental
constraints).  If DM is captured
by the Sun only through inelastic scattering (iDM), this could lead to
a non-negligible fraction of DM annihilating outside of the
Sun's surface.  For models in which iDM annihilates to CREs, an observable
flux at energies above a few tens of GeV could be produced.

During its first year of operation, the LAT has collected a
substantial number of CRE events, which has allowed a precise measurement
of the energy spectrum over a broad
energy range from a few $\units{GeV}$
up to $1\units{TeV}$~\cite{Abdo:2009zk,Ackermann:2010ij}.
Furthermore, a directional analysis of the high-energy CRE
events was performed in the Galactic reference
frame~\cite{Ackermann:2010ip}, and showed no evidence of anisotropies.
Since the Sun is moving with respect to the Galactic reference frame,
the previously-reported absence of anisotropies in the CRE flux observed in the Galactic
frame does not necessarily imply a negative result.
Therefore a dedicated analysis to search for flux variations on CRE correlated 
with the Sun's direction has been perfomed~\cite{Ajello:2011dq}, using two complementary 
analysis approaches: (i) flux asymmetry analysis and (ii) comparison of the solar flux 
with the isotropic flux.

The first approach compares the CRE flux from the Sun with the flux evaluated in a 
sky position opposite to that of the Sun. The second approach used in this analysis is based on the
event-shuffling technique employed in Ref.~\citen{Ackermann:2010ip}, which was used
to build a simulated sample of isotropic CREs starting from the real events. 
Simulated events are built by randomly coupling the arrival times and the arrival directions 
(in local instrument coordinates) of real events. Both analyses did not provide 
any evidence of a CRE  excess from the Sun, so statistical upper limits have been set.

Constraints on DM model parameters by comparing upper limits on solar CRE fluxes to the predicted
fluxes of the two DM annihilation scenarios considered in Ref.~\citen{Schuster:2009fc} 
have been calculated: (1) capture of DM particles
by the Sun via elastic scattering interactions and
subsequent annihilation to $e^{\pm}$ through an
intermediate state $\phi$, and (2) capture of DM
particles by the Sun via inelastic scattering interactions
and subsequent annihilation of the captured DM particles
outside the Sun directly to $e^{\pm}$.
In Ref.~\citen{Ackermann:2010ip} the cases of solar capture by
spin-independent scattering and spin-dependent scattering have been considered separately.

Here we consider only the case of $e^{\pm}$ from annihilation of
DM particles captured by the Sun but with orbits which take them
outside the surface of the Sun. In a standard WIMP scenario,
DM particles captured by the Sun via elastic scattering quickly
undergo subsequent scatterings which cause them to settle to the
core, and hence the fraction of captured DM particles outside the
surface of the Sun at any given time is negligible \cite{Sivertsson:2009nx}.

For a DM particle $\chi$ to scatter inelastically off a nucleon
$N$ via the process $\chi + N \rightarrow \chi^{\star} + N$,
the DM must have energy $E \ge \delta(1+m_{\chi}/m_{N})$,
where $\delta=m_{\chi^{\star}}-m_{\chi}$.
Particles captured by the Sun by inelastic scattering typically
lose enough energy after only a few interactions to prevent
further energy loss by scattering.  If the elastic scattering
cross-section is sufficiently small
($\sigma_{n} \lesssim 10^{-47} \units{cm^{2}}$, e.g.,
Ref.~\citen{Schuster:2009fc}), the captured particles will be
unable to thermalize and settle to the core, and instead
will remain on relatively large orbits.  As a result, the
density of captured DM particles outside the Sun may not be
negligible in an iDM scenario, and the annihilation of
those particles to $e^{\pm}$ could thus produce an observable flux of
CREs from the direction of the Sun.  While it is not necessary
for DM to annihilate primarily to $e^{\pm}$ in order to explain the
direct detection results (since direct detection experiments are not
sensitive to the dominant annihilation channels), leptophilic
iDM is strongly motivated since it could provide a consistent
interpretation of multiple data sets \cite{Finkbeiner:2007kk,
ArkaniHamed:2008qn,Batell:2009zp,Cholis:2009va}.

\begin{figure}[!ht]
\begin{center}
\includegraphics[width=0.9\textwidth]{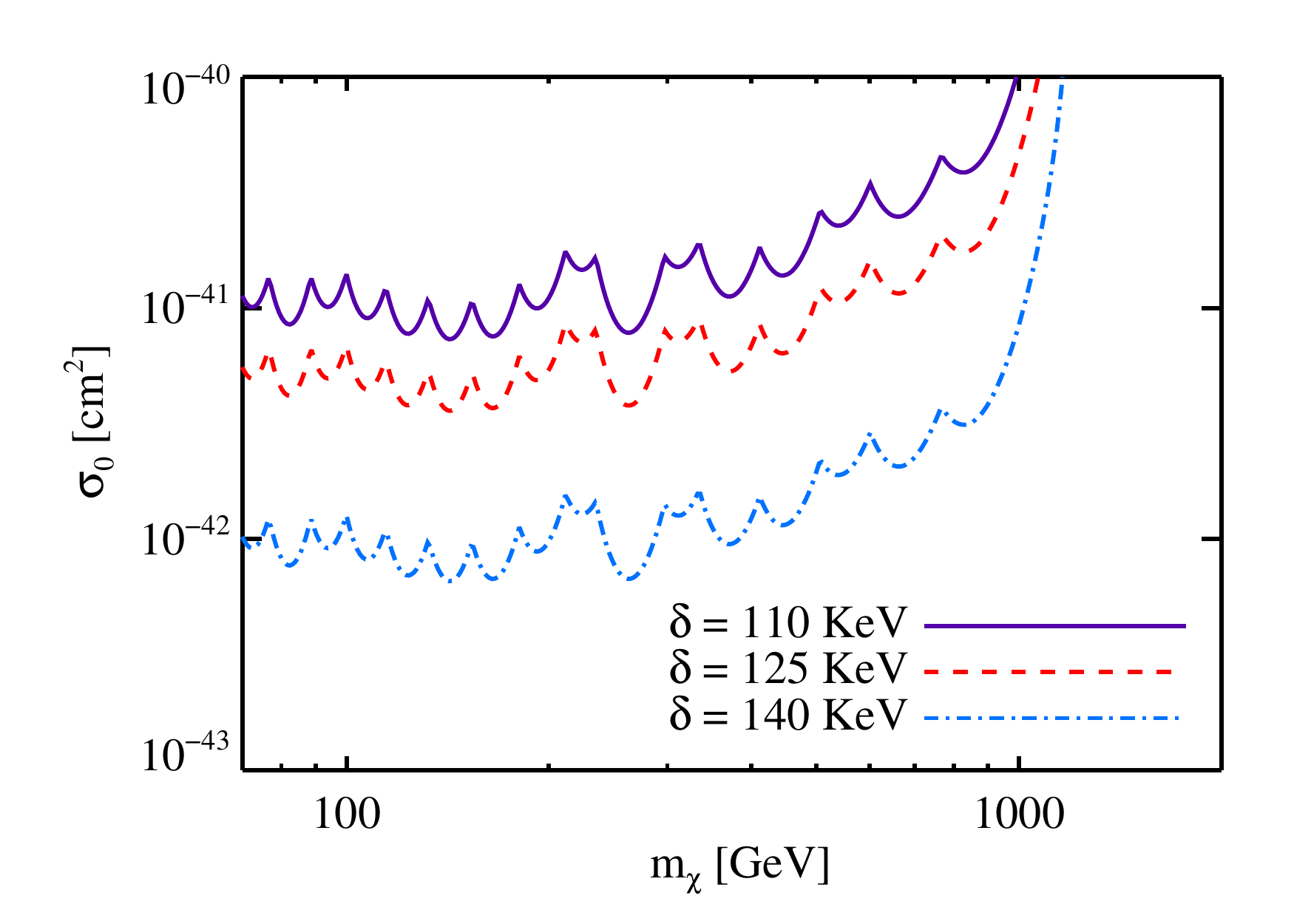}
\end{center}
\caption{Constraints on iDM model parameters for three values
of the mass splitting $\delta$. Models above the curves produce
a solar CRE flux that exceeds the $95\%$ CL flux upper limit
for a $30\degrees$ ROI centered on the Sun in one or more energy bins. Reprinted from Ref.~\citen{Ajello:2011dq}.
\label{fig:idmlim}}
\end{figure}

Fig.~\ref{fig:idmlim} shows the constraints from the solar
CRE flux upper limits on iDM models in the
$m_{\chi}$-$\sigma_{0}$ parameter space for three values
of $\delta$.
These limits exclude the regions of parameter space compatible
with the results of DAMA/LIBRA and CDMS (in addition to several
other direct detection experiments) as determined
by Ref.~\citen{Ahmed:2010hw} for $\delta=120\units{keV}$, for the range of masses
accessible to our analysis ($m_\chi \gtrsim 70\units{GeV}$),
assuming the dominant annihilation channel is
$e^{\pm}$.  Models consistent with both DAMA/LIBRA and CDMS at 90\% CL
exist for values of $\delta$ ranging from $\sim 85\units{keV}$ to
$\sim 135\units{keV}$~\cite{Ahmed:2010hw};
for masses from 70~GeV to 250 GeV the range of
allowed scattering cross-sections is from $\sigma_{0} \sim 10^{-40} \units{cm^{2}}$
to $\sigma_{0} \sim 10^{-39}$ cm$^{2}$~\cite{Chang:2008gd}.
Although the uncertainties in the calculation of the
DM fluxes in this scenario are significant, we emphasize that
constraining $\sigma_{0} \lesssim 10^{-40}\units{cm^{2}}$ is sufficient
to exclude the cross-sections of models consistent with
both data sets.  The bounds we derive exclude
the relevant cross-sections by 1-2 orders of magnitude,
and hence we conclude that the parameter space of models
preferred by DAMA/LIBRA can be confidently ruled out for
$m_\chi \gtrsim 70\units{GeV}$ for annihilation
to $e^{\pm}$ despite the uncertainties in the flux
calculation.
Recently the XENON100 collaboration~\cite{Aprile:2012nq} sets the most stringent limit based on 225 
days of operating time on the spin-independent elastic WIMP-nucleon scattering cross section 
for WIMP masses above 8 $GeV/c^2$, with a minimum of $2 \times 10^{-45}~cm^2$ at 
55 $GeV/c^2$ and 90\% confidence level.

This analysis constrains DM models in which the
primary annihilation channel is to $e^{\pm}$.  We
emphasize that although other annihilation channels can
be probed by gamma-ray \cite{Atkins:2004qr,Batell:2009zp,Schuster:2009au}
or neutrino \cite{Nussinov:2009ft,Menon:2009qj,Schuster:2009au}
measurements, the upper limits on solar CRE fluxes provide
a uniquely strong constraint on the $e^{\pm}$ final state,
which is inaccessible to neutrino telescopes since
no neutrinos are produced for this annihilation channel.

\section{Conclusions}
The Fermi LAT team has looked for indirect DM signals using a wide variety of methods, and since
no signals have been detected, strong constraints have been set. Fermi turned five years 
in orbit on June, 2013, and it is definitely living up to its expectations in terms of 
scientific results delivered to the community. The Fermi-LAT Collaboration will provide 
an improved event reconstruction (Pass 8)~\cite{Atwood:2013rka}, that will improve the 
potential of the LAT instrument.

\section*{Acknowledgments}
The Fermi LAT Collaboration acknowledges generous ongoing support
from a number of agencies and institutes that have supported both the
development and the operation of the LAT as well as scientific data analysis.
These include the National Aeronautics and Space Administration and the
Department of Energy in the United States, the Commissariat \`a l'Energie Atomique
and the Centre National de la Recherche Scientifique / Institut National de Physique
Nucl\'eaire et de Physique des Particules in France, the Agenzia Spaziale Italiana
and the Istituto Nazionale di Fisica Nucleare in Italy, the Ministry of Education,
Culture, Sports, Science and Technology (MEXT), High Energy Accelerator Research
Organization (KEK) and Japan Aerospace Exploration Agency (JAXA) in Japan, and
the K.~A.~Wallenberg Foundation, the Swedish Research Council and the
Swedish National Space Board in Sweden.

Additional support for science analysis during the operations phase is gratefully
acknowledged from the Istituto Nazionale di Astrofisica in Italy and the Centre National d'\'Etudes Spatiales in France.


\end{document}